\documentclass[a4paper,11pt]{article}

\usepackage{jheppub} 

\usepackage[T1]{fontenc} 

\title{\boldmath Constraints on the Universal Varying Yukawa Couplings: from SM-like to Fermiophobic}

\author[a,b]{Tianjun Li,}
\author[c]{Xia Wan,}
\author[a]{You-kai Wang,}
\author[c,d]{and Shou-hua Zhu}

\affiliation[a]{State Key Laboratory of Theoretical Physics
and Kavli Institute for Theoretical Physics China (KITPC),
Institute of Theoretical Physics, Chinese Academy of Sciences,\\
Beijing 100190, P. R. China}
\affiliation[b]{George P. and Cynthia W. Mitchell Institute for
             Fundamental Physics,  Texas A$\&$M University,\\
             College Station,  TX 77843,  USA}
\affiliation[c]{Institute of Theoretical Physics $\&$ State Key Laboratory of
Nuclear Physics and Technology, Peking University,\\ Beijing 100871,
China}
\affiliation[d]{Center for High Energy Physics, Peking University,\\
Beijing 100871, China}

\emailAdd{tli@itp.ac.cn}
\emailAdd{xia.wan@pku.edu.cn}
\emailAdd{wangyk@itp.ac.cn}
\emailAdd{shzhu@pku.edu.cn}

\abstract{Varying the Standard Model (SM) fermion Yukawa couplings universally
by a generic positive scale factor ($F_{Yu}$), we study the phenomenological fit to the
current available experimental results for the Higgs boson search at hadron colliders.
We point out that the Higgs production cross section and its decay
branching ratio to $\gamma\gamma$ can be varied oppositely by $F_{Yu}$ to make
their product almost invariant. Thus, our scenario and the SM Higgs are
indistinguishable in the inclusive $H\to \gamma\gamma$ channel.
The current measurements on direct Yukawa coupling strength in  the
$H\to b\bar{b}/\tau\tau$ channel are not precise enough to
fix the scale factor $F_{Yu}$. The most promising is the vector-boson-fusion
channel in which the CMS has already observed possible suppression
effect on the Yukawa couplings. Further more, the global $\chi^2$ fit of the experimental data
can get the optimal value by introducing a suppression factor $F_{Yu}\sim1/2$ on the SM Yukawa couplings.}

\begin{document}
\maketitle
\flushbottom

\section{Introduction}
\label{Introduction}

Even though the Standard Model (SM) has achieved an impressive phenomenological
success, the Higgs mechanism, which breaks the electroweak gauge symmetry and
gives masses to the electroweak gauge bosons, has not
been confirmed. The Higgs mechanism also gives masses to the SM
 fermions through the Yukawa couplings, which are equal to the fermion masses
times the inverse of the vacuum expectation
value (VEV) of the Higgs field. However, the Yukawa couplings do not
provide any explanation for the observed large hierarchy in the fermion masses.

Recently, both the CMS and ATLAS Collaborations have released a series of results about
the Higgs searches in $\gamma \gamma$, WW, ZZ, $b\bar{b}$ and $\tau\tau$ decay
modes~\cite{CMS-combined, ATLAS-combined, CMS-moriond, ATLAS-moriond}. The CMS
experiment excluded the 127.5-600~GeV at 95\% confidence level while observed
an excess with a global significance of $1.6\sigma$ after look-else-where effect
for a Higgs boson mass hypothesis of 125 GeV in the $\gamma\gamma$
mode~\cite{CMS-moriond}. The ATLAS experiment gives a consistent result
of observing an excess with a significance of $1.5\sigma$ at 126.5~GeV
in the $\gamma\gamma$ mode\cite{ATLAS-moriond}.
ATLAS also see small excesses in $ZZ\to 4l$ channel~\cite{ATLAS-moriond}
at 125~GeV while CMS shows some slight excesses
in $ZZ\to 4l$ channel at 119.5GeV and 320GeV~\cite{CMS-moriond}.
Moreover, Tevatron gives a CDF and D0 combination result of
observing an excess of 2.2$\sigma$ in WH/ZH and $H\to b\bar{b}$
channels~\cite{Tevatron-moriond}. Besides confirming or excluding
these hints experimentally, the urgent problem is to check whether this
125~GeV possible Higgs boson is SM-like or belong to the other SM extensions.

One kind of new physics scenario is the so-called fermiophobic
Higgs model~\cite{theory_fermiophobic_1, theory_fermiophobic_2, theory_fermiophobic_3, theory_fermiophobic_4, theory_fermiophobic_5, theory_fermiophobic_6, theory_fermiophobic_7,Akeroyd:1998ui_theory_fermiophobic, theory_fermiophobic_8, theory_fermiophobic_10, fermiophobic-125},
in which the Higgs boson does not couple to the SM fermions while
 its couplings to the gauge bosons are
the same as the SM. The SM fermions can acquire masses by other mechanisms
rather than the Yukawa couplings. At the LHC, the fermiophobic Higgs production
cross section is much lower than the SM, but its branching
ratio to diphoton increases, thus the experimental observable
cross section $\sigma(pp\to H+x)\cdot \mbox{Br}(H\to \gamma\gamma$)
is slightly changed.
The detailed numbers are shown in table~\ref{table-fermiophobic}.
Recently, both CMS and ATLAS have performed searches for the fermiophobic Higgs boson~\cite{CMS-combined,cms-fermiophobic,ATLAS-fermiophobic}.
For CMS, in Ref.~\cite{CMS-combined}, by combining different decay channels, the fermiophobic
Higgs boson is excluded in the mass range 110-192GeV at $95\%$ CL.
While, analysis in diphoton decay mode with lepton-tagged or
dijet-tagged \cite{cms-fermiophobic} shows there is a small excess of fermiophobic Higgs
with $1.2\sigma$ global significance around 126~GeV. ATLAS has also observed an excess
of fermiophobic Higgs at 125.5~GeV, with a global significance of $1.6\sigma$ in diphoton decay mode~\cite{ATLAS-fermiophobic}.

\begin{table}[tbp]
\centering
\begin{tabular}{|lccc|}
\hline
 Channels & SM & FP & $Ratio(FP/SM)$ \\
\hline
$\sigma(pp\to H+x)$    & 17.50pb & 2.19pb & 12.51\%\\
Br($H\to \gamma\gamma$) & 0.23\% & 1.54\% & 6.70\\
Br($H\to W^+ W^-$) &21.63\% &86.89\%  & 4.02 \\
Br($H\to Z Z$)  & 2.65\% & 10.85\% & 4.10\\
Br($H\to b\bar{b}$)  & 57.54\%  & 0 & 0 \\
Br($H\to \tau\tau$)  & 6.30\% & 0  & 0 \\
\hline
$\sigma(pp\to H+x)\cdot \mbox{Br}(H\to \gamma\gamma$) & 0.04pb& 0.034pb &  0.84\\
$\sigma(pp\to H+x)\cdot \mbox{Br}(H\to W^+ W^-$)  & 3.79pb & 1.90pb &0.50 \\
$\sigma(pp\to H+x)\cdot \mbox{Br}(H\to Z Z$)  &  0.46pb& 0.24pb & 0.51\\
$\sigma(pp\to H+x)\cdot \mbox{Br}(H\to b\bar{b}$)  & 10.70pb & 0 &0 \\
$\sigma(pp\to H+x)\cdot \mbox{Br}(H\to \tau\tau$)  & 1.10pb  & 0  & 0 \\
\hline
\end{tabular}
\caption{\label{table-fermiophobic}Production cross section and decay branching ratios of SM and
fermiophobic Higgs for the mass of 125~GeV at 7~TeV $pp$ collision .}
\end{table}

In the Standard Model, fermions and gauge bosons get their masses from
Yukawa couplings and Higgs kinetic term, respectively. The Higgs
boson, which was initially introduced to have a nonzero VEV,
provides the SM fermion masses via different Yukawa couplings.
 Theoretically, there are quite a few new physics models that have modified
Yukawa couplings. Phenomenologically,
we just introduce a generic positive factor $F_{Yu}$ on the SM Yukawa couplings
for all the SM fermions~\footnote{Generically, a negative factor can also be valid. Such as discussed in~\cite{Espinosa:2012ir,Giardino:2012ww}. We will compare positive and negative $F_{Yu}$ in the following sections.}. The standard model is recovered when $F_{Yu}=1$, and
it is the pure fermiophobic case when $F_{Yu}=0$. As the coupling strength
is modified, the SM fermions should get their mass by other scenarios or Higgs field.
The simplest way is to put a $SU(2)_L\times U(1)_Y$ breaking term
$m_q \bar{q}_L q_R$~\cite{Quigg:2009vq}.
Other possible solution can be realized by, as suggested in Ref.~\cite{Quigg:2009vq},
the ``Extended technicolor''. To be model independent, We will not
go though in the details of such models here.

In this paper,  we consider the universal varying SM fermion Yukawa couplings
by a generic positive scale factor. We discuss the phenomenological fit to the
current available hadron collider Higgs boson search experimental results,
and find that the optimal fit gives the $F_{Yu}\sim 0.5$.
 Interestingly, the Higgs production cross section and its decay
branching ratio to $\gamma\gamma$ can be varied oppositely so that
their product is almost invariant. Thus, the inclusive $H\to \gamma\gamma$
decay channel information is not enough to confirm the SM
Higgs boson. The current experimental results on direct Yukawa coupling
channels $H\to b\bar{b}/\tau\tau$ are not precise enough to
determine the scale factor. The most promising channel is the vector boson fusion
since the CMS Collaboration has shown that the SM fermion couplings are
 probably smaller than the SM.

The paper is organized as follows: in section~\ref{model}, we describe the
models and related formulas for Higgs production and decay channels
at the LHC. Section~\ref{numerical} presents the numerical results and comparison
with experimental results for the 125~GeV Higgs. Section~\ref{conclusion} is the conclusion.
Because the ATLAS and CMS collaborations at the LHC announced the discovery of
a new Higgs-like boson before the publication of the paper, we add an update supplementary
analysis in section~\ref{supplement}.

\section{The model\label{model}}

As we only introduce a modified factor on the Yukawa coupling,
the Higgs couplings to the gauge bosons are unchanged. At the
LHC, the dominant SM Higgs production process is the gluon-gluon
fusion channel, via a heavy quark triangle vertex diagram.
This process can be seriously changed by the modified Yukawa factor.
The leading order production cross section can be described as~\cite{Djouadi:2005gi}:
\begin{equation}
\sigma_{LO}^{gg}(pp\to H+x)=\sigma_0^H \tau_H \frac{d {\cal{L}}^{gg}}{d \tau_H}
\ \ \mbox{with} \ \frac{d {\cal{L}}^{gg}}{d \tau}=\int^1_\tau \frac{d x}{x} g(x, \mu_F^2) g(\tau/x, \mu_F^2)~,~\,
\end{equation}
where $\tau_H=M_H^2/s$ with $s$ being the invariant collider energy squared,
$g(x, \mu_F^2)$ is the parton-distribution-function and
$\mu_F$ is the factorization scale. The parton level cross section $\sigma_0^H$ is
\begin{equation}
\sigma_0^H=\frac{G_F\alpha_s^2(\mu_R^2)}{288\sqrt{2}\pi}\left|\frac{3}{4}\sum\limits_Q F_{Yu} A_{1/2}^H(\tau_Q)\right|^2~,~\,
\end{equation}
where $F_{Yu}$ is the modified Yukawa coupling factor, $G_F$ is the Fermi
constant, $\alpha_s$ is the strong coupling constant,
 $\mu_R$ is the renormalization scale, and
$\tau_Q={M_H^2}/{4m_Q^2}$. The quark amplitude $A_{1/2}^H(\tau)$ is
\begin{equation}
A_{1/2}^H(\tau)=2[\tau+(\tau-1)f(\tau)]\tau^{-2}~.~\,
\label{A}
\end{equation}
The function $f(\tau)$ is,
\begin{equation}
f(\tau)=\Big\{
\begin{array}{lr}
\arcsin^2\sqrt{\tau} & \tau\leq1\\
-\frac{1}{4}[\log\frac{1+\sqrt{1-\tau^{-1}}}{1-\sqrt{1-\tau^{-1}}}-i\pi]^2 & \tau>1 \\
\end{array}~.~\,
\label{ftau}
\end{equation}
The K-factor for the gluon-gluon fusion process comes mainly from high order
QCD corrections, which is not affected by the modified Yukawa coupling factor.

Other main production processes include the vector-boson-fusion (VBF) process,
the associated production with $W$/$Z$ bosons ($WH/ZH$ process),
and the associated Higgs production with  heavy top quarks ($t\bar{t}H$ process).
The VBF process and $WH/ZH$ processes  are unaffected by the
modified Yukawa coupling as there is no Yukawa coupling in these processes.
They are dominant production channels in the fermiophobic Higgs model
in which the gluon-gluon fusion process disappears.
For the $t\bar{t}H$ process which depends on the Yukawa
coupling, due to its small portion, the impact from modified Yukawa coupling
can be neglected safely although we really consider it in our calculation.

For Higgs decays, the 125~GeV SM Higgs decays mainly to $b\bar{b}$.
The decay width is highly suppressed when $F_{Yu}$ approaches
to zero. The Born level decay width $\Gamma(H\to f\bar{f})$ can be calculated
by:
\begin{equation}
\Gamma(H\to f\bar{f})=\frac{G_F N_c}{4\sqrt{2}\pi}F_{Yu}^2 M_H m_f^2\beta_f^3~,~\,
\end{equation}
where $N_c=3$, $m_f$ is the fermion mass, and $\beta=(1-4m_f^2/M_H^2)^{1/2}$.

Although the branching ratio is small, the $H\to \gamma \gamma$ process is one
of the most promising channels for a low mass Higgs search at LHC due to its clean
 and simple final state topology, as well as the precise photon reconstruction on CMS and ATLAS detectors.
The $H\to \gamma \gamma$ process is a combination of both $W$ and top quark triangle
loops, whose decay width can be described as:
\begin{equation}
\Gamma(H\to \gamma\gamma)=\frac{G_F\alpha^2 M_H^3}{128\sqrt{2}\pi^3}
\left|\sum\limits_f N_c Q_f^2 F_{Yu} A_{1/2}^H(\tau_f)+A_1^H(\tau_W)\right|^2~,~\,
\label{equation-hrr}
\end{equation}
where $Q_f$ is the quark's charge,
$A_{1/2}^H(\tau)$ and $A_{1}^H(\tau)$ are form factors for the spin $1/2$ and spin $1$ particles
respectively, and $\tau_i = M_H^2/4M_i^2$ with $i = f, W$. $A_{1/2}^H(\tau)$ is given in eq.~(\ref{A}).
\begin{equation}
A_1^H(\tau)=-[2\tau^2+3\tau+3(2\tau-1)f(\tau)]\tau^{-2}~,~\,
\label{equation-A-W}
\end{equation}
$f(\tau)$ is given in eq.~(\ref{ftau}).
As $A_1^H(\tau_W)$ has an opposite minus sign to $A_{1/2}^H(\tau_f)$, the conjugation term between fermions
and $W$ bosons is suppressed when $0<F_{Yu}<1$. Thus, the $\Gamma(H\to \gamma\gamma)$ will
increase when $F_{Yu}$ approaches zero.

In this paper, we do not consider the case when $F_{Yu}<0$. In most cases,
the cross sections are proportional to $|F_{Yu}|^2$.
The only difference for a negative $F_{Yu}$ is happened in the $H\to \gamma\gamma$ decay,
as shown in equation~(\ref{equation-hrr}). For a positive $F_{Yu}<1$,
$\Gamma(H\to \gamma\gamma)$ is increased by reducing the minus conjugation term between $W$ and fermion loops;
For a negative $F_{Yu}$, the conjugation term
between $W$ and fermion loop is positive, so the $\Gamma(H\to \gamma\gamma)$ is increased.
Related discussion can be found in~\cite{Espinosa:2012ir,Giardino:2012ww}.

The $H\to WW^{\ast}/ZZ^{\ast}$ processes are not
affected by modifying the Yukawa factor.

By suppressing Yukawa couplings, the dominant production process $gg\to H$ can be reduced.
Meanwhile, the SM dominant decay mode, $\Gamma(H\to b\bar{b})$ becomes small so the total decay
width decrease. $\Gamma(H\to \gamma\gamma)$ can be increased by
reducing the cancelation between $W$ and top quark loops.
The branching ratio $\mbox{Br}(H\to \gamma\gamma)$ is increased. In all,
by introducing the modified Yukawa factor $0<F_{Yu}<1$,
the decrease range of $\sigma(gg\to H)$ can be nearly the same as the
increase range of $\mbox{Br}(H\to \gamma\gamma)$ to make their product almost
stable, which has been shown explicitly in table~\ref{table-fermiophobic} for $F_{Yu}=0$.

\section{Numerical results\label{numerical}}

In this Section, we shows the comparisons between the theoretical results after
introducing the Yukawa factor $F_{Yu}$ and the experimental constraints.
The SM Higgs production cross sections and their corresponding errors are taken
from the state-of-art estimations by CERN Higgs Cross section Working
Group~\cite{cern-lhc-twiki}.
The Higgs decay branching ratios are calculated by using the HDECAY package~\cite{hdecay}
with modifying the Yukawa factor according to formulae described in section~\ref{model}.
The SM parameters are taken as $G_F=1.16637\times10^{-5}~\mbox{GeV}^{-2}$, $\alpha_s(m_Z)=0.119$,
and $m_t=172.5$GeV. The Higgs mass is fixed at 125~GeV by default.

\begin{figure}[htbp]
\begin{center}
\includegraphics[width=0.5\textwidth]
{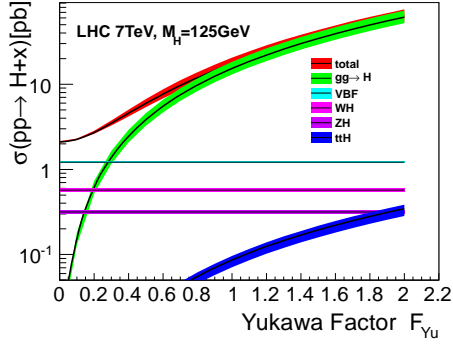}
\end{center}
\caption{\label{producingCS}Higgs production cross section
versus Yukawa factor $F_{Yu}$ at the LHC.}
\end{figure}

Figure~\ref{producingCS} shows the production cross section of Higgs boson versus
the modified Yukawa factor $F_{Yu}$. It can be seen that the
gluon-gluon fusion process is greatly impacted by this factor. The total cross
section is mainly composed of gluon-gluon fusion when $F_{Yu}>0.3$. When
$F_{Yu}<0.2$, $\sigma(pp\to H+x)$ is mainly contributed by VBF, $WH$ and $ZH$ processes,
which are independent of the variation of $F_{Yu}$. The $\sigma(pp\to ZH+X)$
is about half of the $\sigma(pp\to WH+X)$, which is mainly caused by
$W/Z$ different couplings to quarks.
The $t\bar{t}H$ channel's contribution is less than $1\%$ even though
its value can be comparable with $ZH$ process when $F_{Yu}\sim 2$.

\begin{figure}[htbp]
\begin{center}
\includegraphics[width=0.49\textwidth]
{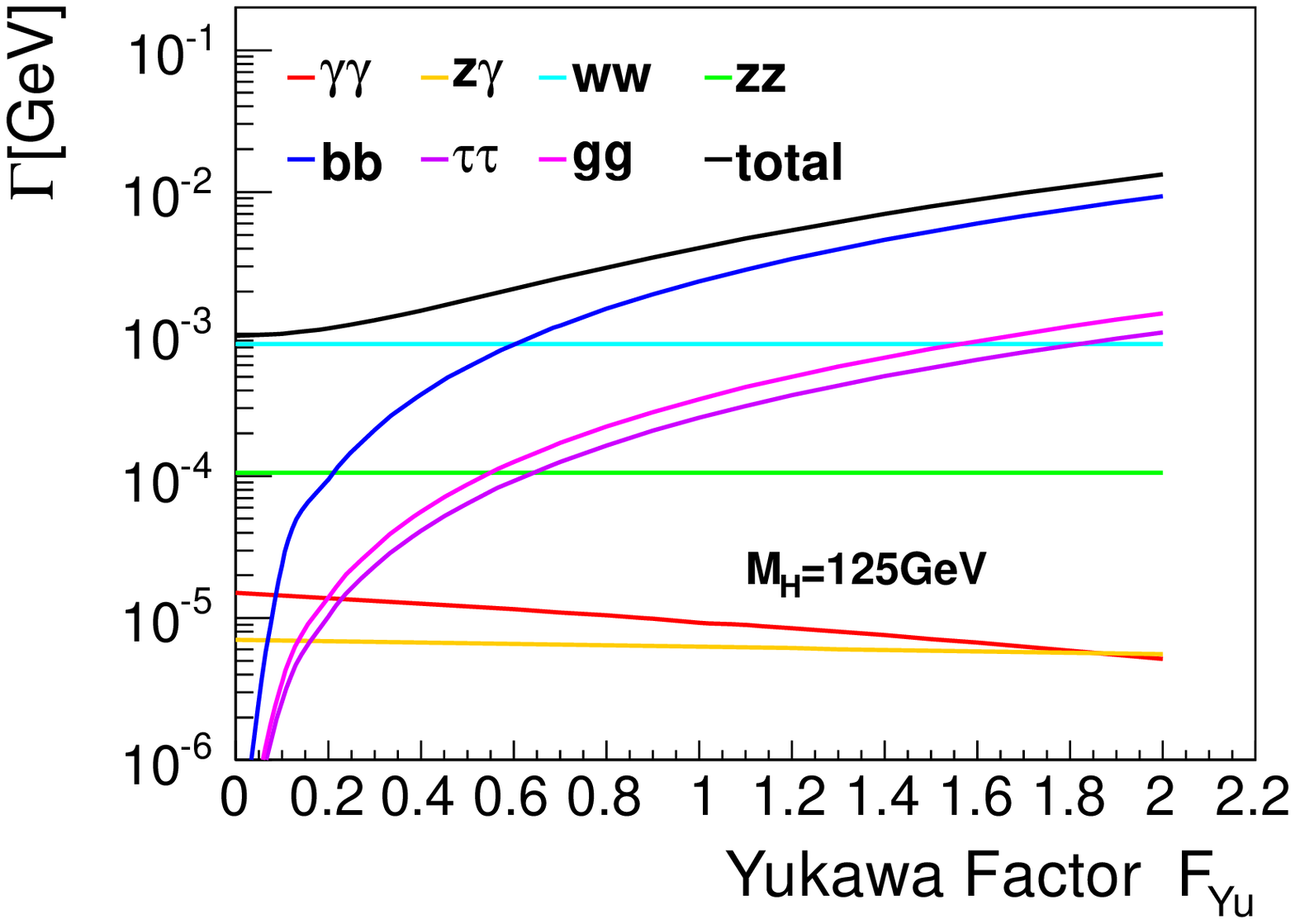}
\includegraphics[width=0.49\textwidth]
{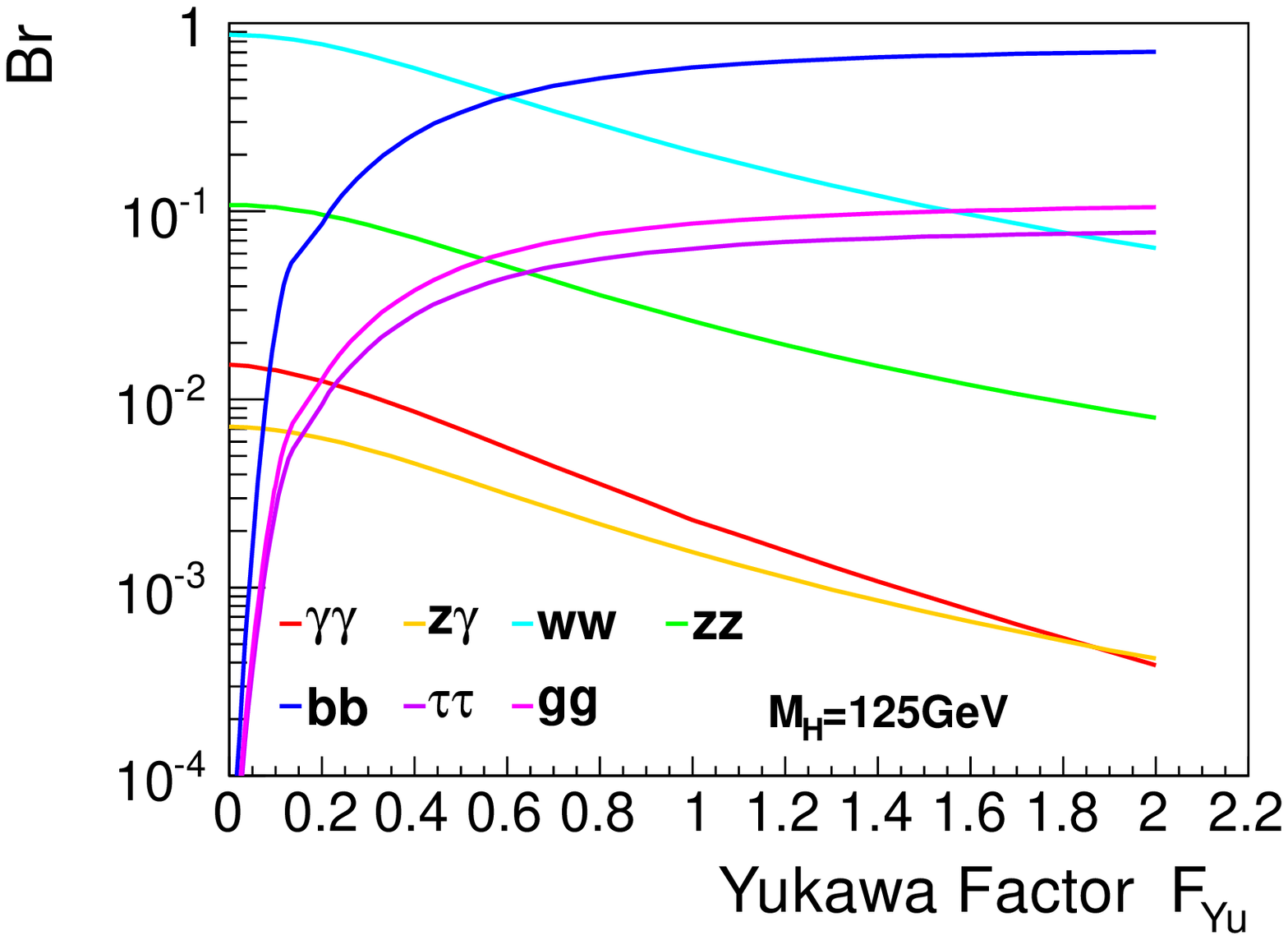}
\end{center}
\caption{\label{decay}Higgs decay widths and branching ratios versus Yukawa factor $F_{Yu}$.}
\end{figure}

Figure~\ref{decay} shows the decay widths and the branching ratios for different
Higgs decay modes versus $F_{Yu}$. These channels can be divided into
3 categories.
\begin{enumerate}
\item The gauge boson decay mode $H\to WW^{\ast}/ZZ^{\ast}$ whose partial decay widths are insensitive
to $F_{Yu}$.
\item The $H\to \gamma\gamma$ and $Z\gamma$ channels which are partly
impacted by $F_{Yu}$. These channels are contributed by both $W$ and
heavy fermion loops. Only the fermion-Higgs vertex in the triangle diagrams are affected.
\item The $H\to b\bar{b}$, $\tau\tau$ and $gg$ channels which are sensitively
affected. Their decay widths are proportional to $F_{Yu}^2$.
\end{enumerate}

\begin{figure}[htbp]
\begin{center}
\includegraphics[width=0.49\textwidth]
{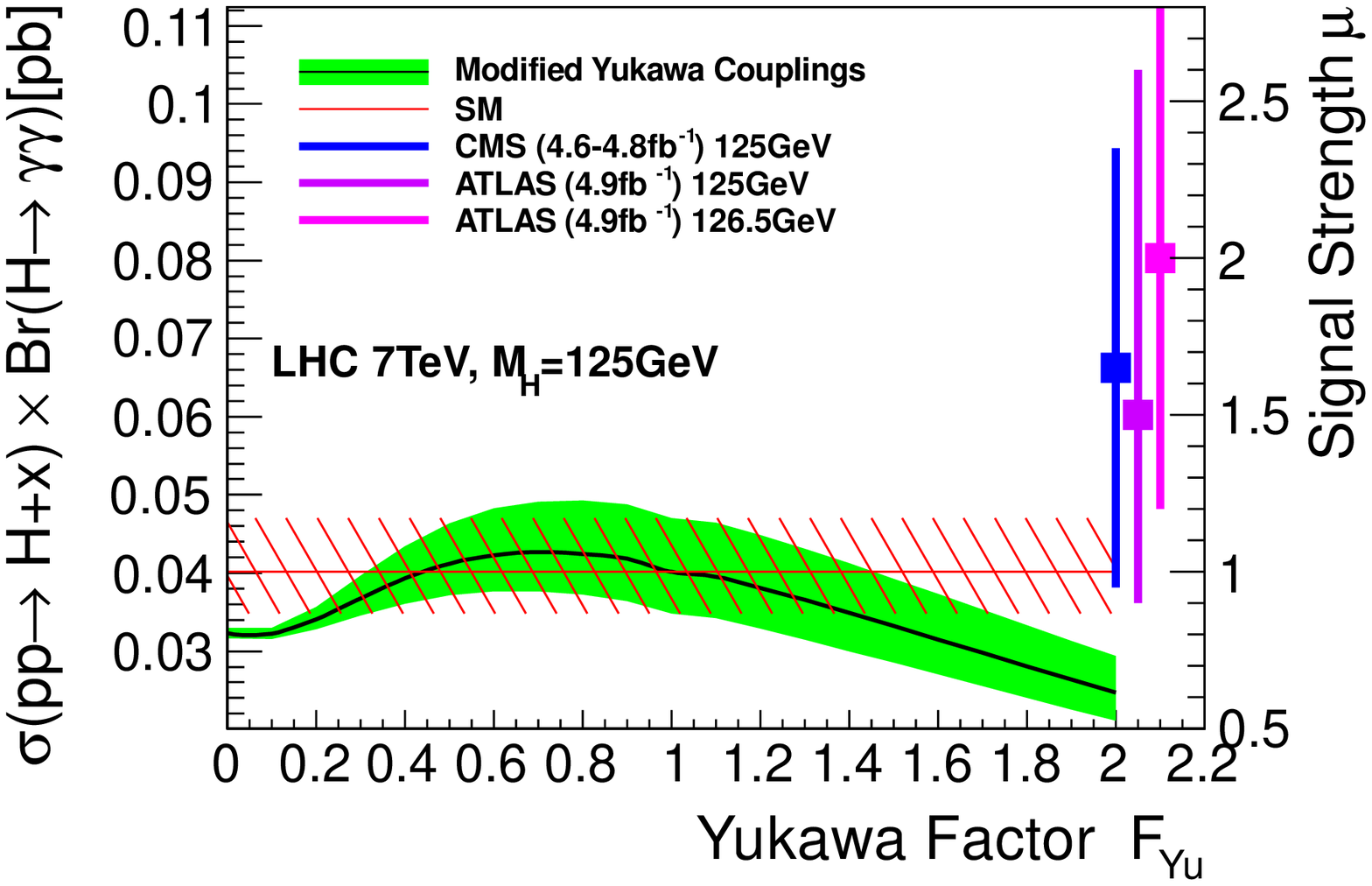}\\
\includegraphics[width=0.49\textwidth]
{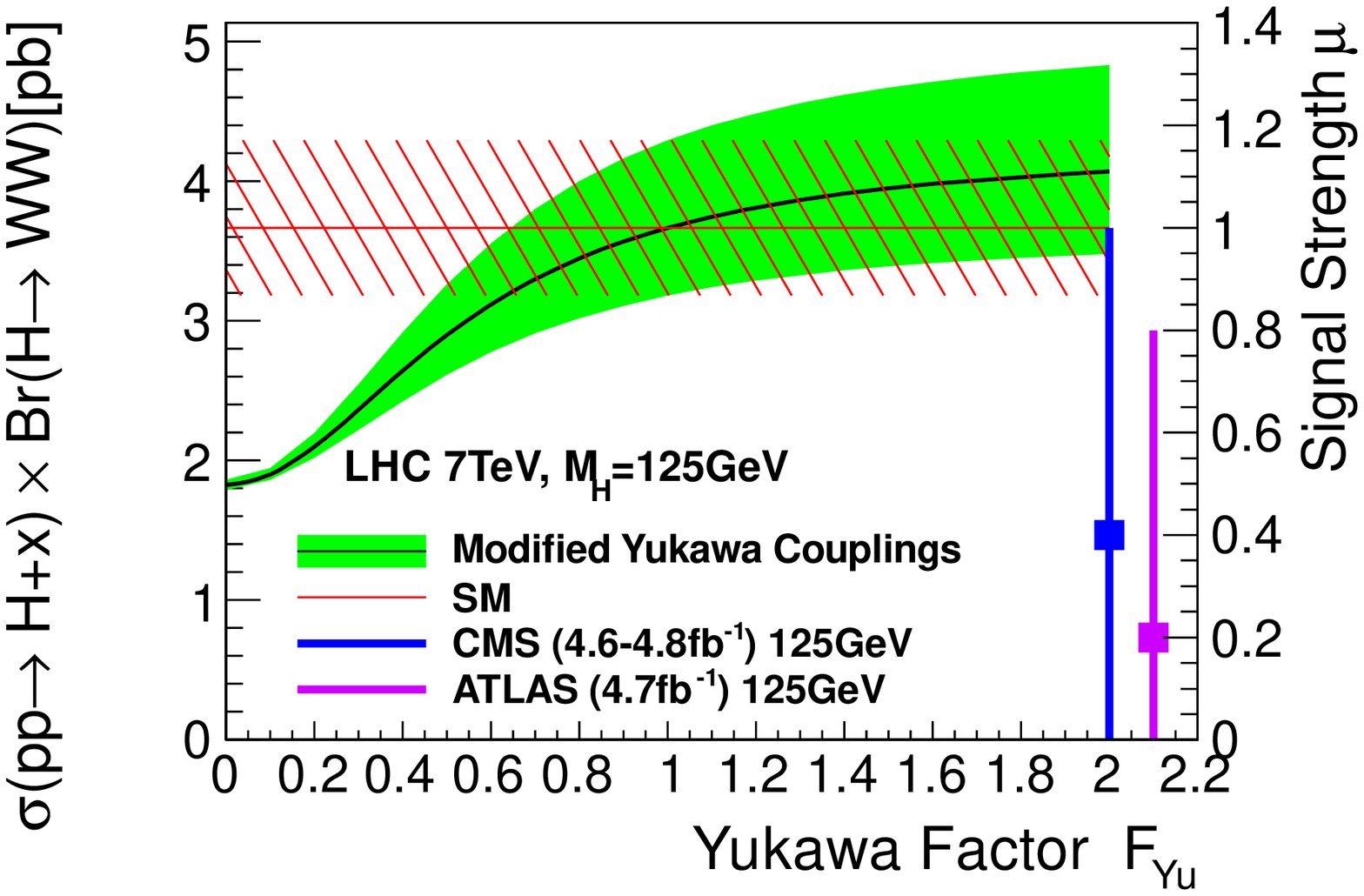}
\includegraphics[width=0.49\textwidth]
{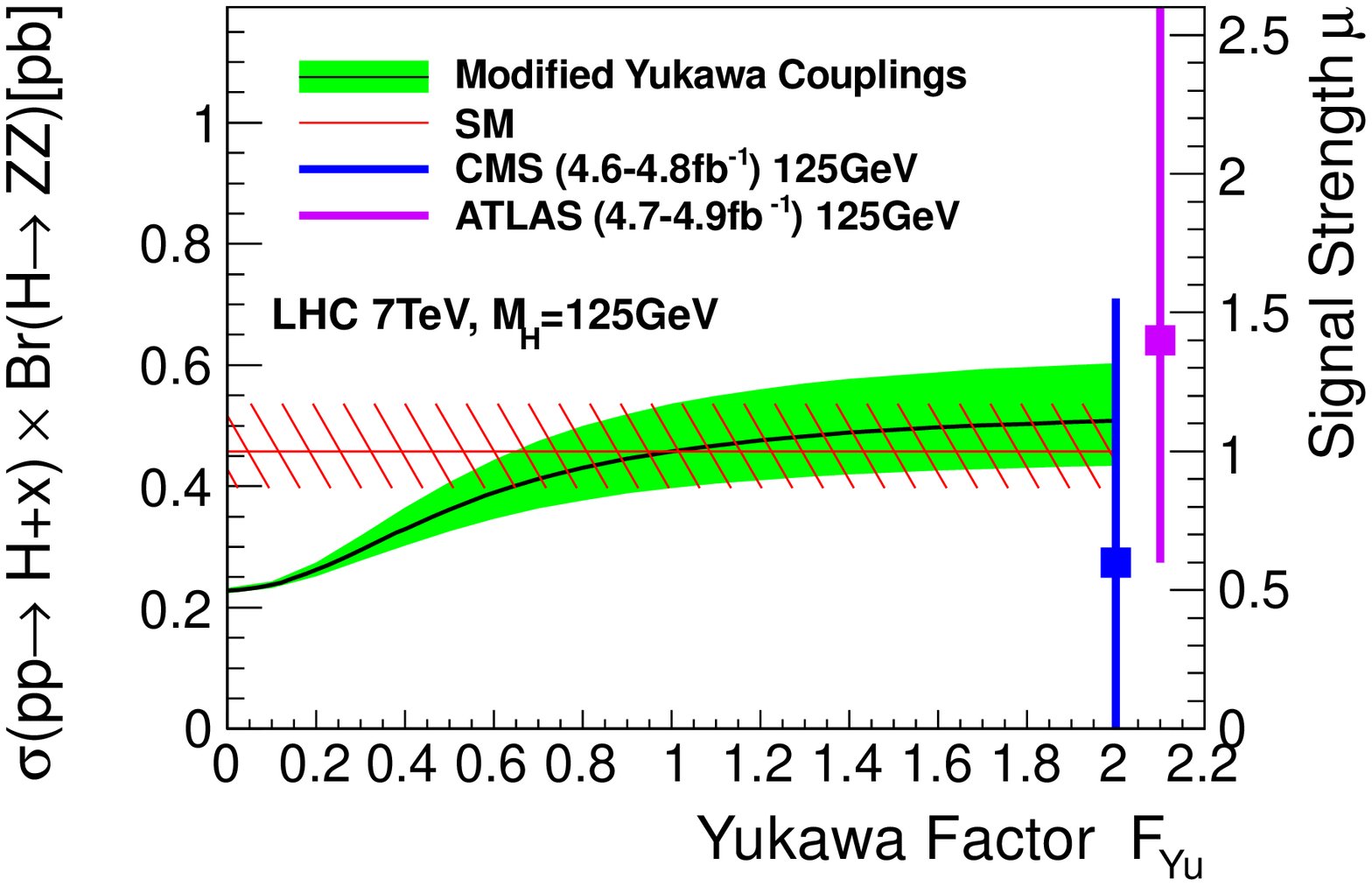}
\end{center}
\caption{\label{CSgaugebosons} Higgs observable cross sections ($\sigma\times \text{Br}$) versus
Yukawa factor $F_{Yu}$ for $H\to\gamma\gamma, WW^{\ast}, ZZ^{\ast}$ channels (black curve). The green hatched
region presents the uncertainties. The SM value ($F_{Yu}=1$) and its uncertainties are highlighted
with the red line plus red hatched region for a convenient comparison. The square points with error bars
show the CMS and ATLAS experimental results of the best-fit signal strength value $\mu = \sigma/\sigma_{SM}$
as labeled on the right side Y-axis. Same conventions applied in the following plots.
  }
\end{figure}

\begin{figure}[htbp]
\begin{center}
\includegraphics[width=0.49\textwidth]
{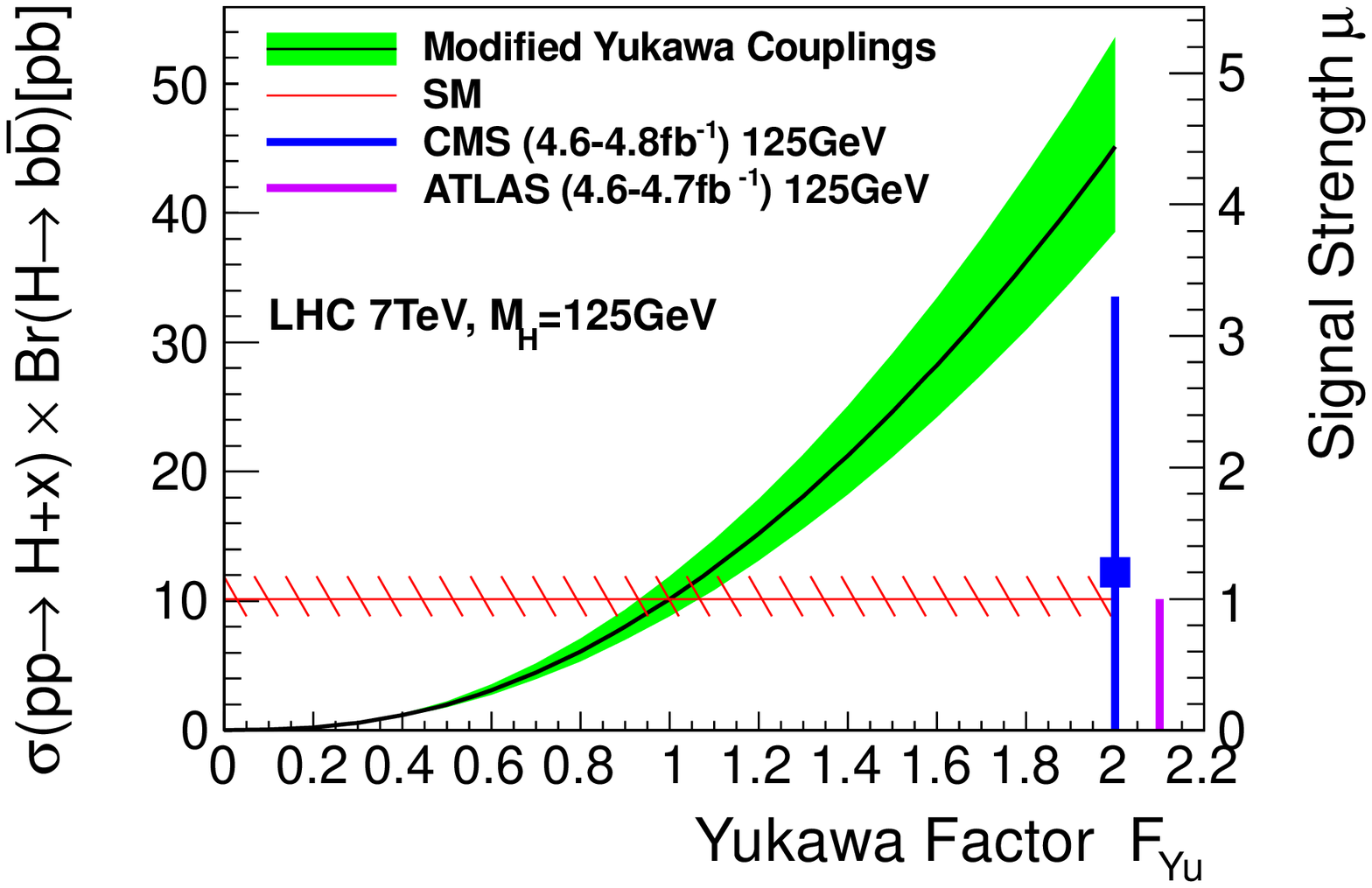}
\includegraphics[width=0.49\textwidth]
{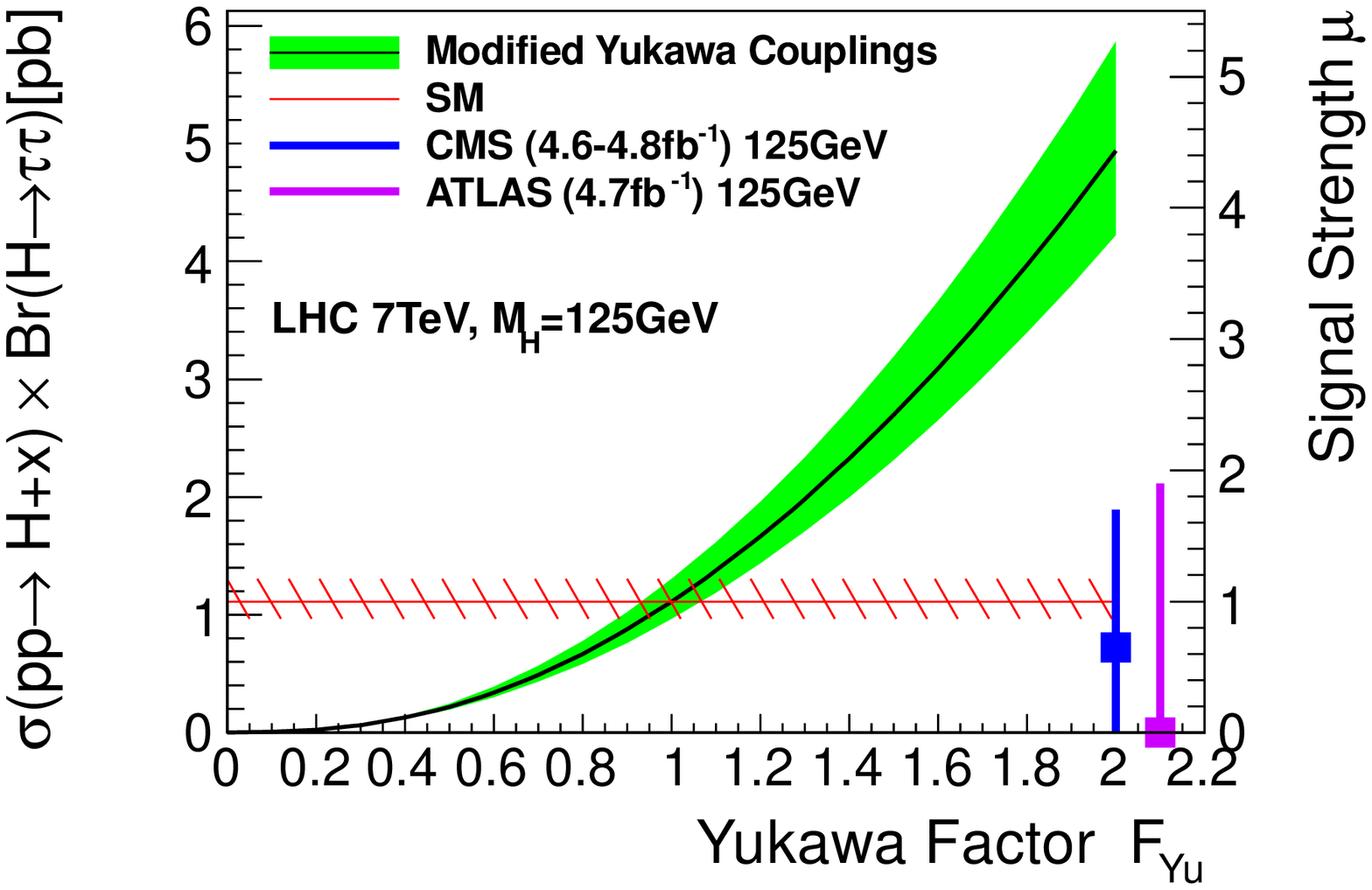}
\end{center}
\caption{\label{CSff} Higgs observable cross sections ($\sigma\times \text{Br}$) versus
Yukawa factor $F_{Yu}$ for $H\to b\bar{b}/\tau\tau$ channels.}
\end{figure}

\begin{figure}[htbp]
\begin{center}
\includegraphics[width=0.49\textwidth]
{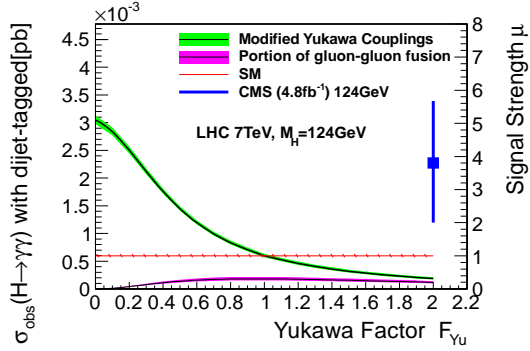}
\end{center}
\caption{\label{VBF} $H\to \gamma\gamma$ observable cross sections with dijet-tagged versus Yukawa factor $F_{Yu}$. The cross sections are obtained by $\sigma_{VBF}(pp\to H+x)\text{Br}(H\to \gamma\gamma)\cdot15\%+\sigma_{gg-fusion}(pp\to H+x)\text{Br}(H\to \gamma\gamma)\cdot0.5\%$, in which $15\%(0.5\%)$ is the efficiency for VBF(gg-fusion) process\cite{Chatrchyan:2012tw}. The black curve with purple hatched region shows the gg-fusion portion. }
\end{figure}

\begin{figure}[htbp]
\begin{center}
\includegraphics[width=0.49\textwidth]
{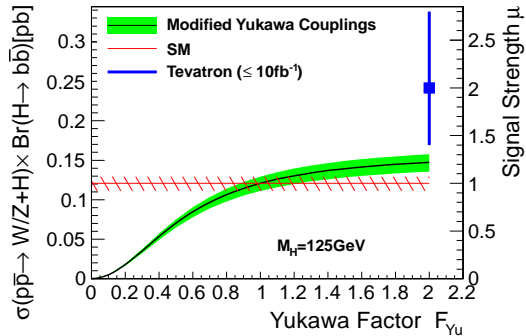}
\end{center}
\caption{\label{tevatron-bb} The $WH/ZH$, $H\to b\bar{b}$ process at Tevatron with $M_H=125$~GeV.}
\end{figure}

Comparisons of the model predictions and the experimental results
are given in Figure~\ref{CSgaugebosons},~\ref{CSff},~\ref{VBF} and \ref{tevatron-bb}.
For the modified Yukawa couplings, we have only considered uncertainties
from the production cross section which are linear combination of the
QCD scale and PDF+ $\alpha_{s}$ uncertainties~\cite{cern-lhc-twiki}.
For the hadron collider experimental results, they are taken from the best-fit
of signal strength $\mu=\sigma/\sigma_{SM}$ figures, as shown in table~\ref{table-expdata}.
By assuming the same selection efficiency as the SM signal,
they can be compared directly to the production cross sections in our scenario,
with normalized by SM production cross sections.
Although curves and SM predicted values in our figures are drawn mostly with $M_H=125$~GeV,
the experimental data of the excess peaks can have small shifts.
For example, the ATLAS excess peak in $H\to \gamma\gamma$ channel is located at 126.5~GeV,
whose signal strength is sensitive to this mass shift. To be clear,
we show both $M_H=125$~GeV and 126.5~GeV $\mu$ values in table~\ref{table-expdata}
and figure~\ref{CSgaugebosons}. In principle, curve should be drawn
with $M_H=126.5$~GeV to compare with the same Higgs mass experimental $\mu$ data.
However, for the inclusive Higgs production with diphoton decay,
the SM observation cross sections are $\sigma \cdot \text{Br}(H\to\gamma\gamma)=17.50\times2.29\times10^{-3} =0.0401$~pb for $M_H=125$~GeV and $\sigma \cdot \text{Br}(H\to\gamma\gamma)=17.07\times2.29\times10^{-3} =0.0391$~pb for $M_H=126.5$~GeV~\cite{cern-lhc-twiki}.
So the difference between the small mass split can be safely neglected.
Another special case is the CMS $H\to \gamma\gamma$ with dijet-tagged,
which corresponds to $M_H=124$~GeV. Similar as the above explanation, it is a
good approximation to neglect the small difference between $M_H=124$ or $125$~GeV in the comparison.
Expect the two cases discussed above, the signal strength $\mu$ are
insensitive to the small mass shift and they are taken with $M_H=125$~GeV.

\begin{table}[htb]
\centering
\begin{tabular}{|lccc|}
\hline
 &CMS & ATLAS & Tevatron \\
\hline
$H\to \gamma\gamma$ & $1.65^{+0.7}_{-0.7}$\cite{CMS-combined}&
\begin{tabular}{c}
$2.0^{+0.8}_{-0.8}|_{M_H=126.5\text{GeV}}$\cite{ATLAS-combined}\\
$1.5^{+0.9}_{-0.6}$\cite{ATLAS-combined}\\
\end{tabular}
&$\cdots$
\\
$H\to W^+ W^-$ & $0.4^{+0.6}_{-0.6}$\cite{CMS-combined}&$0.2^{+0.6}_{-0.7}$\cite{ATLAS-combined} &$\cdots$\\
$H\to Z Z$  & $0.6^{+0.95}_{-0.6}$\cite{CMS-combined}&$1.4^{+1.2}_{-0.8}$\cite{ATLAS-combined} &$\cdots$\\
$H\to b\bar{b}$  & $1.2^{+2.1}_{-1.9}$\cite{CMS-combined}&$-0.8^{+1.8}_{-1.7}$\cite{ATLAS-combined} &$\cdots$\\
$H\to \tau\tau$  & $0.65^{+1.05}_{-1.3}$\cite{CMS-combined}&$0.0^{+1.9}_{-1.9}$\cite{ATLAS-combined} &$\cdots$\\
$H\to\gamma\gamma$ with dijet-tagged & $3.8^{+2.4}_{-1.8}|_{M_H=124\text{GeV}}$\cite{Chatrchyan:2012tw}&$\cdots$ &$\cdots$ \\
$WH/ZH, H\to b\bar{b}$ &$\cdots$ &$\cdots$ & $2.0^{+0.8}_{-0.6}$\cite{Tevatron-moriond}\\
\hline
\end{tabular}
\caption{\label{table-expdata} A collection of best-fit
of signal strength $\mu=\sigma/\sigma_{SM}$ experimental data. The values are read at $M_H=125$~GeV by default and special cases are labeled otherwise. }
\end{table}

For $H\to \gamma\gamma$ channel with $F_{Yu}$
increases from 0 to 2, the production cross section
increases and $\text{Br}(H\to \gamma\gamma)$ decreases, which makes
$\sigma(pp\to H +X)\cdot\text{Br}(H\to \gamma\gamma)$ change slowly around
the SM predicted value $0.04$pb. Thus, the SM and our modified
Yukawa scenario will be indistinguishable in the inclusive
$H\to \gamma\gamma$ decay mode.

For $H\to WW^{\ast}$ or $ZZ^{\ast}$, they both have very similar
curves on the plots. The 125~GeV Higgs decays at least into one off-shell gauge boson
due to it is under the threshold mass. The only difference is, $\text{Br}(H\to WW^{\ast})$
is about 10 times of $\text{Br}(H\to ZZ^{\ast})$. So their observable cross sections can have about
one order difference. The current $WW^{\ast}/ZZ^{\ast}$ experimental results do not have enough
precision to fix $F_{Yu}$ although $H\to WW^{\ast}$ tends to be smaller than the SM prediction.

The information of the direct Yukawa coupling measurement channels $H\to b\bar{b}$
and $H\to \tau\tau$ at the LHC~\cite{CMS-moriond,ATLAS-moriond}
are shown in figure~\ref{CSff}.
Large Yukawa factor $F_{Yu}$ region ($F_{Yu}>2$) can be excluded by current experimental results,
but the error bar is too large to give a strict constraint on the $F_{Yu}$.

As we have emphasized above,
the pure gauge boson $\gamma\gamma$, $WW^{\ast}$ and $ZZ^{\ast}$ Higgs decay
channels may have difficulties to distinguish fermiophobic or our modified Yukawa
models from the SM. Another complimentary mode is the VBF Higgs
production and its decay into two photons, which makes the main contribution in
Figure~\ref{VBF}(Part of Higgs produced by gg-fusion can also
pass the dijet-tagging selection cuts as shown in the figure).
The advantage of the VBF production channel is, its production
cross section does not change with $F_{Yu}$. The component of
$\sigma_{VBF}(pp\to H+x)\cdot \text{Br}(H\to \gamma\gamma)$ just exhibits the
variation of the branching ratio $\Gamma(H\to \gamma\gamma)$ with $F_{Yu}$. By tagging
the two forward jets, CMS has given the corresponding results~\cite{Chatrchyan:2012tw}
for 124GeV Higgs and there can be
about 1.5$\sigma$ deviation from the SM predicted values. It
agrees well with our scenario when $F_{Yu}\sim 0.3$.
If this deviation can be confirmed in future experiments, it will be the direct evidence
to support a smaller Yukawa factor rather than the SM.

Recently, Tevatron has also released their results on Higgs search~\cite{Tevatron-moriond}.
Broad excess has been observed in the $WH/ZH, H\to b\bar{b}$ channel.
Tevatron data in the table~\ref{table-expdata} is a combination result for $H\to b\bar{b}$,
which is dominantly contributed by $ZH\to ll b\bar{b}$ process~\cite{Tevatron-moriond}.
This channel is important to fix $F_{Yu}$ for our scenario as it has
direct $Hb\bar{b}$ Yukawa coupling. The signal strength $\sigma/\sigma_{SM}$
for 125~GeV Higgs as been shown in figure~\ref{tevatron-bb}. There is no overlap
between our selected $F_{Yu}$ region and the experimental data.

\begin{figure}[htbp]
\begin{center}
\includegraphics[width=0.49\textwidth]
{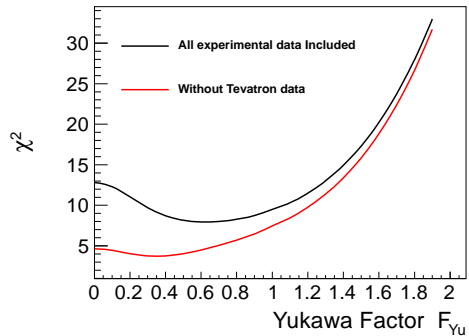}
\end{center}
\caption{\label{chi2} The $\chi^2$ fit to find the optimal Yukawa factor $F_{Yu}$. For the black curve, all experimental data in table~\ref{table-expdata} except ATLAS inclusive $H\to \gamma\gamma$ with $M_H=126.5$~GeV are used. For the red curve, Tevatron experimental data is also excluded. }
\end{figure}

The estimated $F_{Yu}$ parameter regions within one standard deviation for all decay channels
are summarized in table~\ref{table-factor}, which are extracted from
the above figures. For a conservative estimation, value of $F_{Yu}$
can be taken within allowed region by CMS or ATLAS results.
The comparison of the model predictions with the experimental data by each channel
can exhibit the parameter dependence by different detection modes.
For the complimentary research, we also show the $\chi^2$ distribution variate
with $F_{Yu}$ in Figure~\ref{chi2}. The $\chi^2$ is constructed as:
\begin{equation}
\chi^2=\sum\limits_i \left(\frac{\mu_i^{\mbox{the}}-\mu_i^{\mbox{exp}}}{\delta_i}\right)^2,
\label{equation-chi2}
\end{equation}
in which $\mu_i^{\mbox{the}}$ are theoretical predicted signal strength
as the observation cross sections normalized by the SM cross sections,
$\mu_i^{\mbox{exp}}$ are experimental data as shown in the table~\ref{table-expdata}.
$\delta_i$ are experimental errors and are taken as the average values for asymmetric errors.
The optimal $F_{Yu}$ value, which corresponds the $\chi^2_{min}$,
is 0.3(0.6) for Tevatron included(excluded) data sample.
Roughly speaking, a suppression effects with about half of the SM Yukawa coupling strength
can obtain a optimal global fit.

\begin{table}[tbp]
\centering
\begin{tabular}{|lcc|}
\hline
 Collider &Channels & $F_{Yu}$ region  \\
\hline
&$H\to \gamma\gamma$ & 0.25-1.3 \\
&$H\to W^+ W^-$ & 0.0-1.0 \\
LHC&$H\to Z Z$  & 0.0-2.0 \\
&$H\to b\bar{b}$  & 0.0-1.75 \\
&$H\to \tau\tau$  & 0.0-1.35 \\
&$H\to\gamma\gamma$ with dijet-tagged  & 0.0-0.6  \\
\hline
Tevatron &$WH/ZH, H\to b\bar{b}$  & $\cdots$ \\
\hline
\end{tabular}
\caption{\label{table-factor} The allowed $F_{Yu}$ region by different experiment results.  }
\end{table}

\section{Conclusion\label{conclusion}}

LHC has already observed some possible excesses in the Higgs search
at about 125~GeV. For a deeper understanding of the gauge
symmetry breaking and the origins of mass, it is necessary to study
phenomenologically the properties of this possible 125~GeV Higgs. In this
paper, we focus on the Yukawa coupling between Higgs and fermions. By
introducing an effective universal Yukawa coupling scale factor, we show that at the LHC,
the production cross section $\sigma(pp\to H+x)$ and the branching ratio
$\text{Br}(H\to \gamma\gamma)$ can change oppositely to make their product
almost stable. The inclusive $H\to \gamma\gamma$ information is not
capable of distinguishing between the modified Yukawa scenario and the SM.
Due to tagging and reconstructing difficulties, the current LHC $H\to b\bar{b}$ or $\tau\tau$ data
are not precise enough to examine the Yukawa coupling directly. CMS
has observed possible deviation from the SM prediction in the vector-boson-fusion Higgs production
with diphoton decay. This can be explained by a suppressed effective Yukawa coupling.
Our investigation shows that, the global $\chi^2$ fit can get its optimal value
by introducing a suppression factor $F_{Yu}\sim1/2$ on the SM Yukawa couplings.
Limited by the information of the experimental data, $F_{Yu}\sim1/2$ is a very rough estimation.
Future higgs search measurements can crosscheck this estimation and possible model construction
work can be made to suppress the Yukawa couplings.

{\bf Note Added}: While our paper has been completed, we noticed the
paper~\cite{Giardino:2012ww}, which also
studied the Higgs properties from the LHC and Tevatron data.

\section{Supplement}\label{supplement}
Before the publication of this paper, ATLAS and CMS at the LHC announced the
discovery of a new higgs-like boson at a mass near 125~GeV\cite{ATLAS:higgsnote,ATLAS:higgs,CMS:higgsnote,CMS:higgs}.
Therefore, it is necessary
to include the latest experimental results in this research. To keep the self-completeness of
the previous analysis, we make the update analysis in this supplement section.

\begin{figure}[htbp]
\begin{center}
\includegraphics[width=0.5\textwidth]
{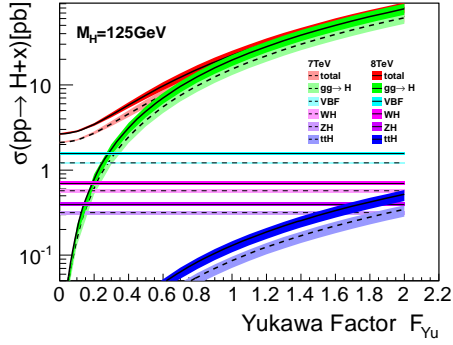}
\end{center}
\caption{\label{producingCS78}Higgs production cross section
versus Yukawa factor $F_{Yu}$ at the LHC for both $\sqrt{s}=$7\&8~TeV.
The solid (dashed) lines with (light) shaded regions show the cross sections at 8~TeV (7~TeV).}
\end{figure}

Figure~\ref{producingCS78} shows the comparison of the Higgs production cross sections versus
$F_{Yu}$ with $\sqrt{s}=$7~TeV and 8~TeV.  The SM Higgs production cross sections
at 8~TeV and their corresponding errors are taken
from the state-of-art estimations by CERN Higgs Cross Section Working
Group~\cite{cern-lhc-twiki8}. For $m_H=125$~GeV, the SM Higgs production cross section at 8~TeV
is about 1.27 times of that at 7~TeV. The Higgs decay width and branching ratio are independent
of the LHC energy as already shown in figure~\ref{decay}.

To compare the model predictions with the experimental results, we collect
the best-fit values of signal strength $\mu$ from the experimental data in
table~\ref{table-expdata78}. To be consistent, all signal strengths $\mu$ of
different channels for ATLAS and CMS are taken as the $7\&8$~TeV combined results,
also shown are their integrated luminosities at $7\&8$~TeV.
The corresponding Higgs mass may have a small (+0.5~GeV or +1.0~GeV) but ignorable difference from 125~GeV.

\begin{table}[htb]
\centering
\begin{tabular}{|lcccc|}
\hline
&\multicolumn{2}{c}{CMS} &\multicolumn{2}{c}{ATLAS} \vline\\
&$\mu$&${\cal L}$(7\&8TeV)&$\mu$&${\cal L}$(7\&8TeV)\\
\hline
$H\to \gamma\gamma$ & $1.54^{+0.48}_{-0.43}$\cite{CMS:higgs}&$5.1\mbox{fb}^{-1}\&5.3\mbox{fb}^{-1}$\cite{CMS:higgs}&
$1.8\pm{0.5}$\cite{ATLAS:higgs}&$4.8\mbox{fb}^{-1}\&5.9\mbox{fb}^{-1}$\cite{ATLAS:higgs}\\
$H\to W^+W^-$ & $0.60^{+0.44}_{-0.41}$\cite{CMS:higgs}&$4.9\mbox{fb}^{-1}\&5.1\mbox{fb}^{-1}$\cite{CMS:higgs}&
$1.3\pm{0.5}$\cite{ATLAS:higgs}&$4.7\mbox{fb}^{-1}\&5.8\mbox{fb}^{-1}$\cite{ATLAS:higgs}\\
$H\to ZZ$ & $0.73^{+0.47}_{-0.37}$\cite{CMS:higgs}&$5.1\mbox{fb}^{-1}\&5.3\mbox{fb}^{-1}$\cite{CMS:higgs}&
$1.4\pm{0.6}$\cite{ATLAS:higgs}&$4.8\mbox{fb}^{-1}\&5.8\mbox{fb}^{-1}$\cite{ATLAS:higgs}\\
$H\to b\bar{b}$ & $0.47^{+0.85}_{-0.71}$\cite{CMS:higgs}&$5.0\mbox{fb}^{-1}\&5.1\mbox{fb}^{-1}$\cite{CMS:higgs}&
$\cdots$&$\cdots$\\
$H\to \tau\tau$ & $0.08^{+0.79}_{-0.76}$\cite{CMS:higgs}&$4.9\mbox{fb}^{-1}\&5.1\mbox{fb}^{-1}$\cite{CMS:higgs}&
$\cdots$&$\cdots$\\
VBF, $H\to \gamma\gamma$ & $2.13^{+1.32}_{-1.09}$\cite{CMS:higgsnote}&$5.1\mbox{fb}^{-1}\&5.3\mbox{fb}^{-1}$\cite{CMS:higgsnote}&
$\cdots$&$\cdots$\\
$VH,H\to b\bar{b}$ & $0.52^{+0.84}_{-0.80}$\cite{CMS:higgsnote}&$5.0\mbox{fb}^{-1}\&5.1\mbox{fb}^{-1}$\cite{CMS:higgsnote}&
$\cdots$&$\cdots$\\
\hline
\end{tabular}
\caption{\label{table-expdata78} A collection of $7\&8$~TeV combined best-fit values
of signal strength $\mu=\sigma/\sigma_{SM}$ from experimental data, together with the integrated luminosities at $7\&8$~TeV.}
\end{table}

\begin{figure}[htbp]
\begin{center}
\includegraphics[width=0.49\textwidth]
{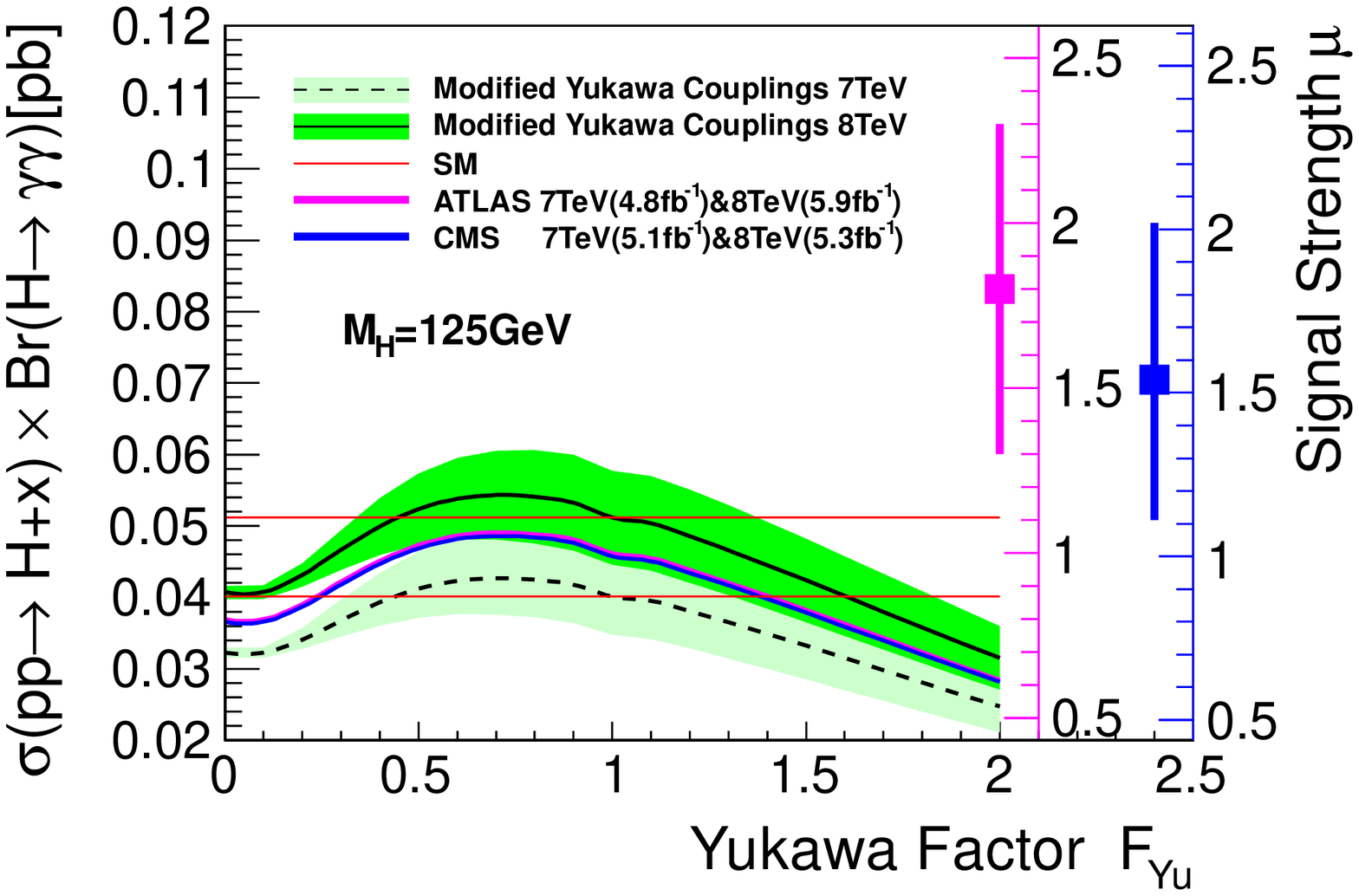}\\
\includegraphics[width=0.49\textwidth]
{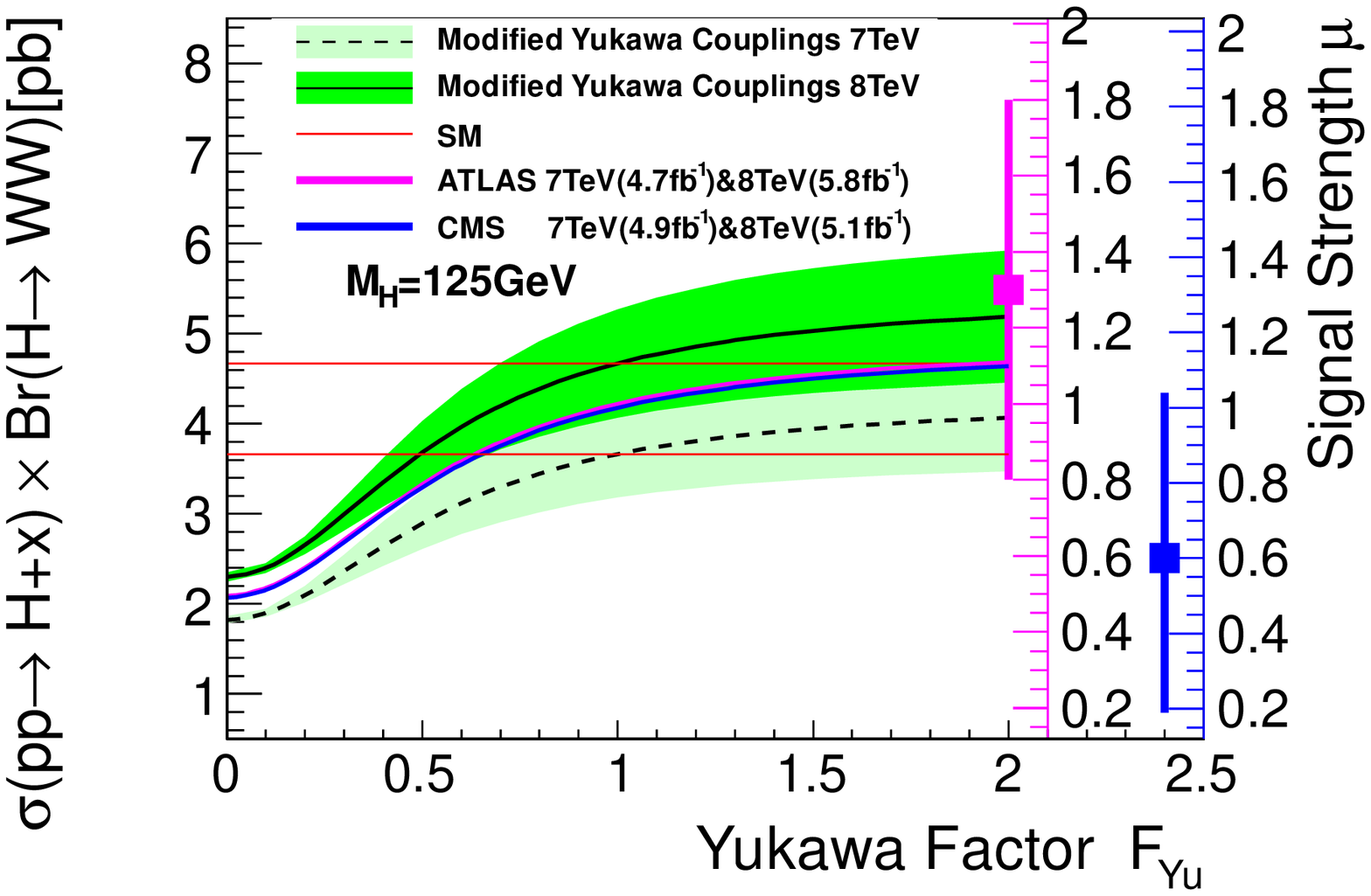}
\includegraphics[width=0.49\textwidth]
{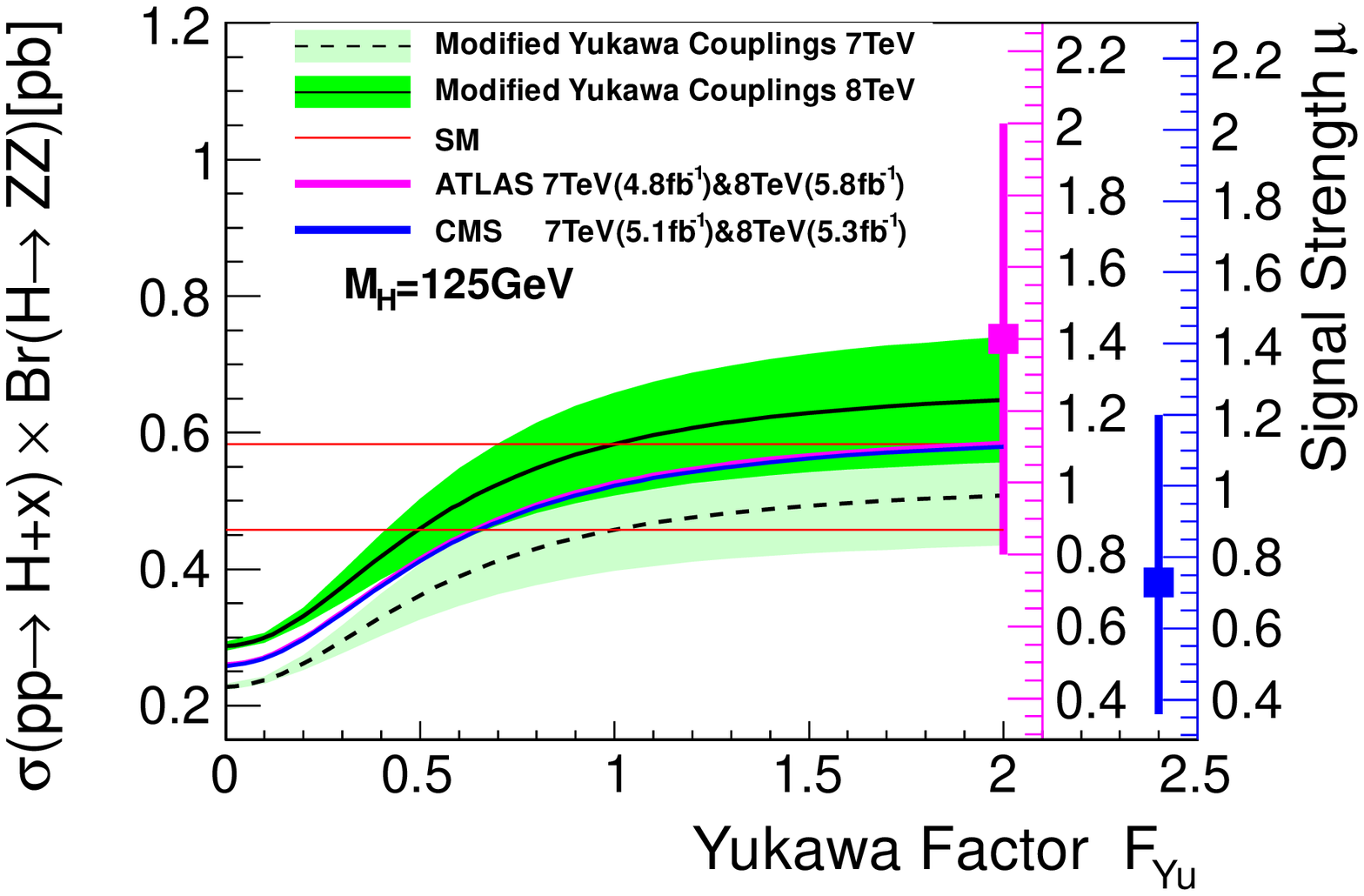}
\end{center}
\caption{\label{CSgaugebosons78} Higgs observable cross sections ($\sigma\times \text{Br}$) versus
Yukawa factor $F_{Yu}$ for $H\to\gamma\gamma, WW^{\ast}, ZZ^{\ast}$ channels (black/dashed curve for 8/7~TeV). The green shaded
region presents the uncertainties. The SM value ($F_{Yu}=1$) are shown
with the red lines for both $7\&8$~TeV cases. The $7\&8$~TeV combined predictions
are drawn in magenta/blue for ATLAS/CMS. The magenta/blue square points with error bars
show the ATLAS/CMS experimental results of the best-fit signal strength value $\mu = \sigma/\sigma_{SM}$, which are labeled on the right side Y-axis individually. Same conventions applied in the following plots.
  }
\end{figure}

\begin{figure}[htbp]
\begin{center}
\includegraphics[width=0.49\textwidth]
{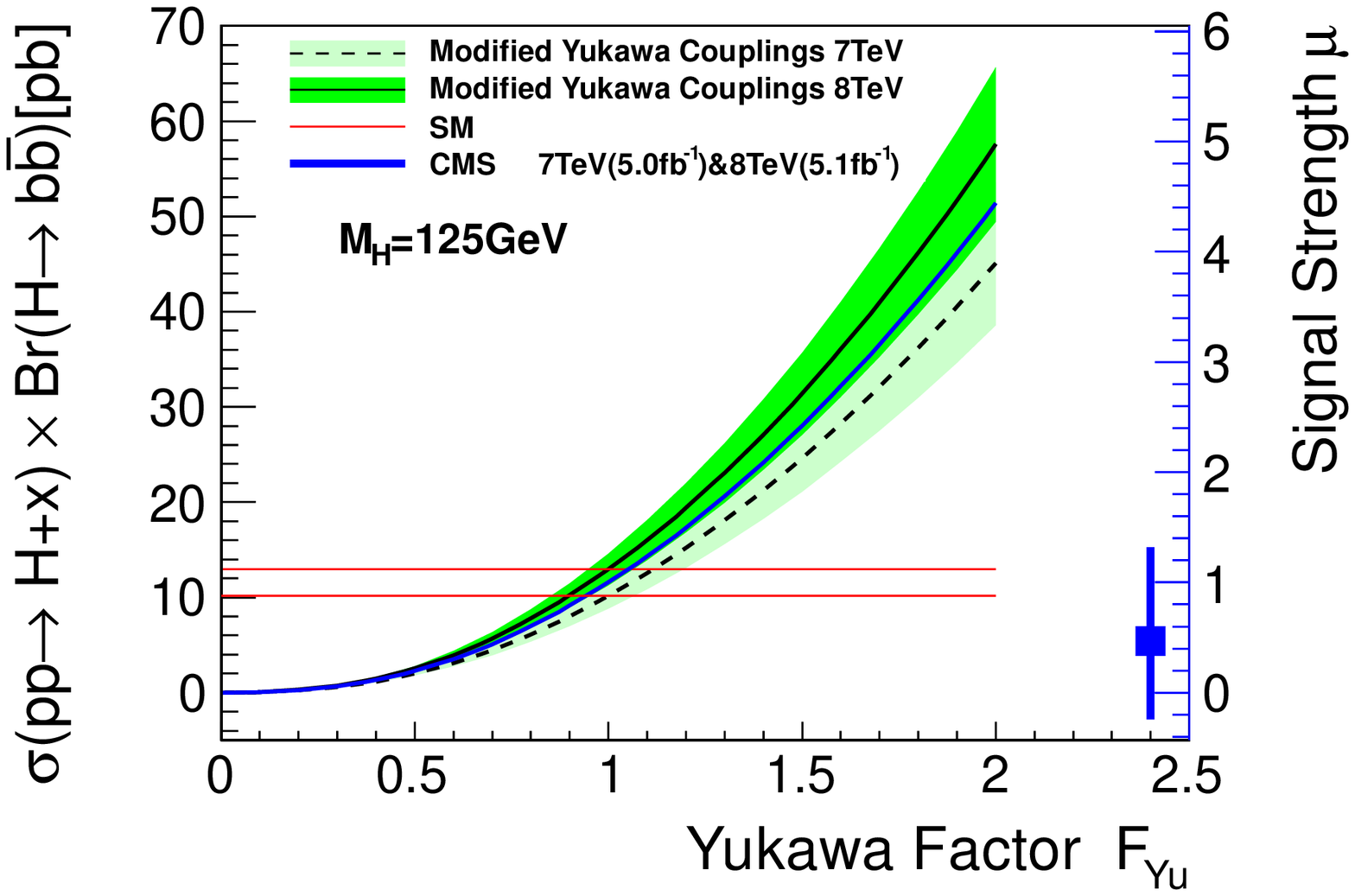}
\includegraphics[width=0.49\textwidth]
{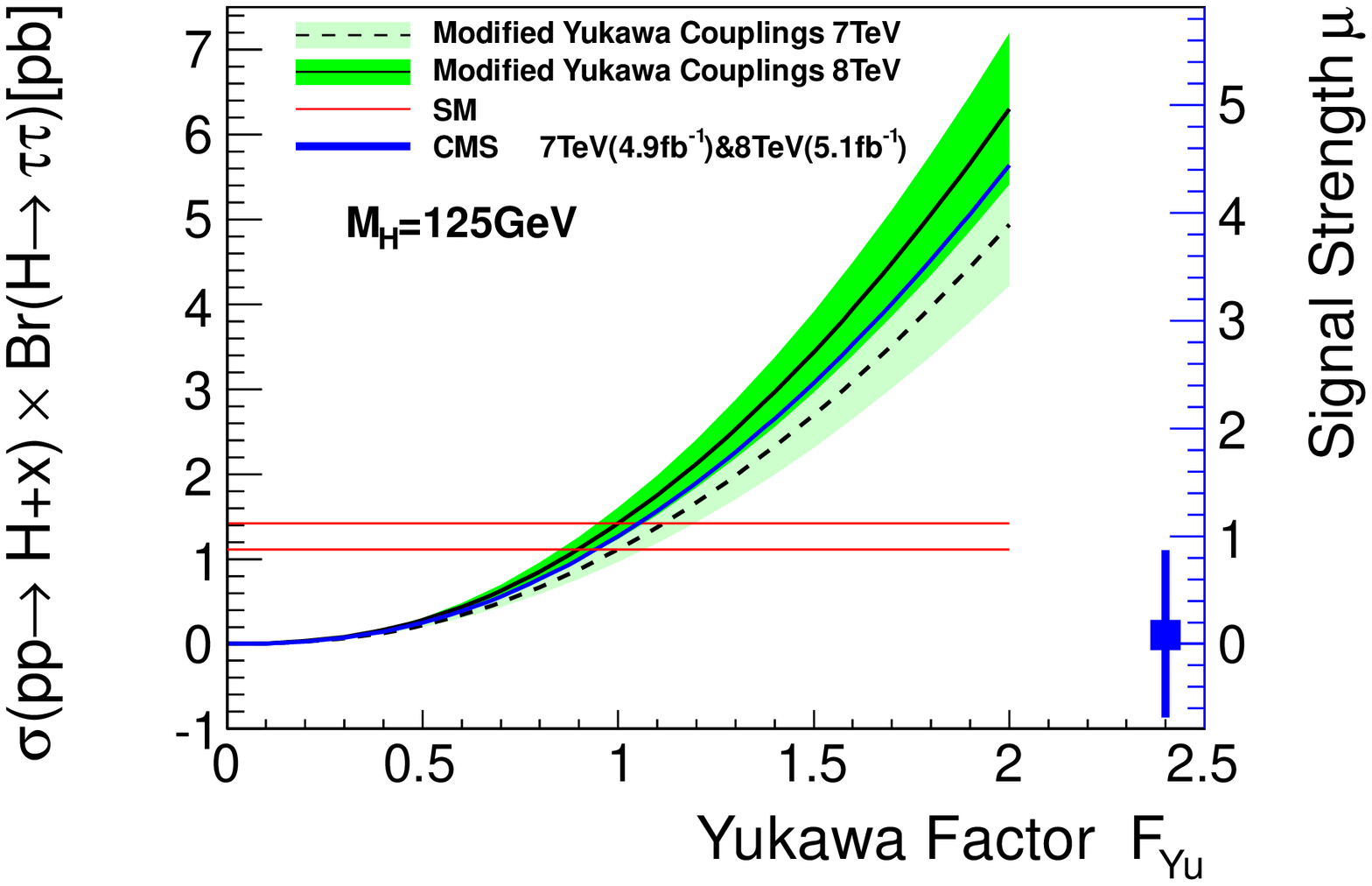}
\end{center}
\caption{\label{CSff78} Higgs observable cross sections ($\sigma\times \text{Br}$) versus
Yukawa factor $F_{Yu}$ for $H\to b\bar{b}/\tau\tau$ channels.}
\end{figure}

\begin{figure}[htbp]
\begin{center}
\includegraphics[width=0.49\textwidth]
{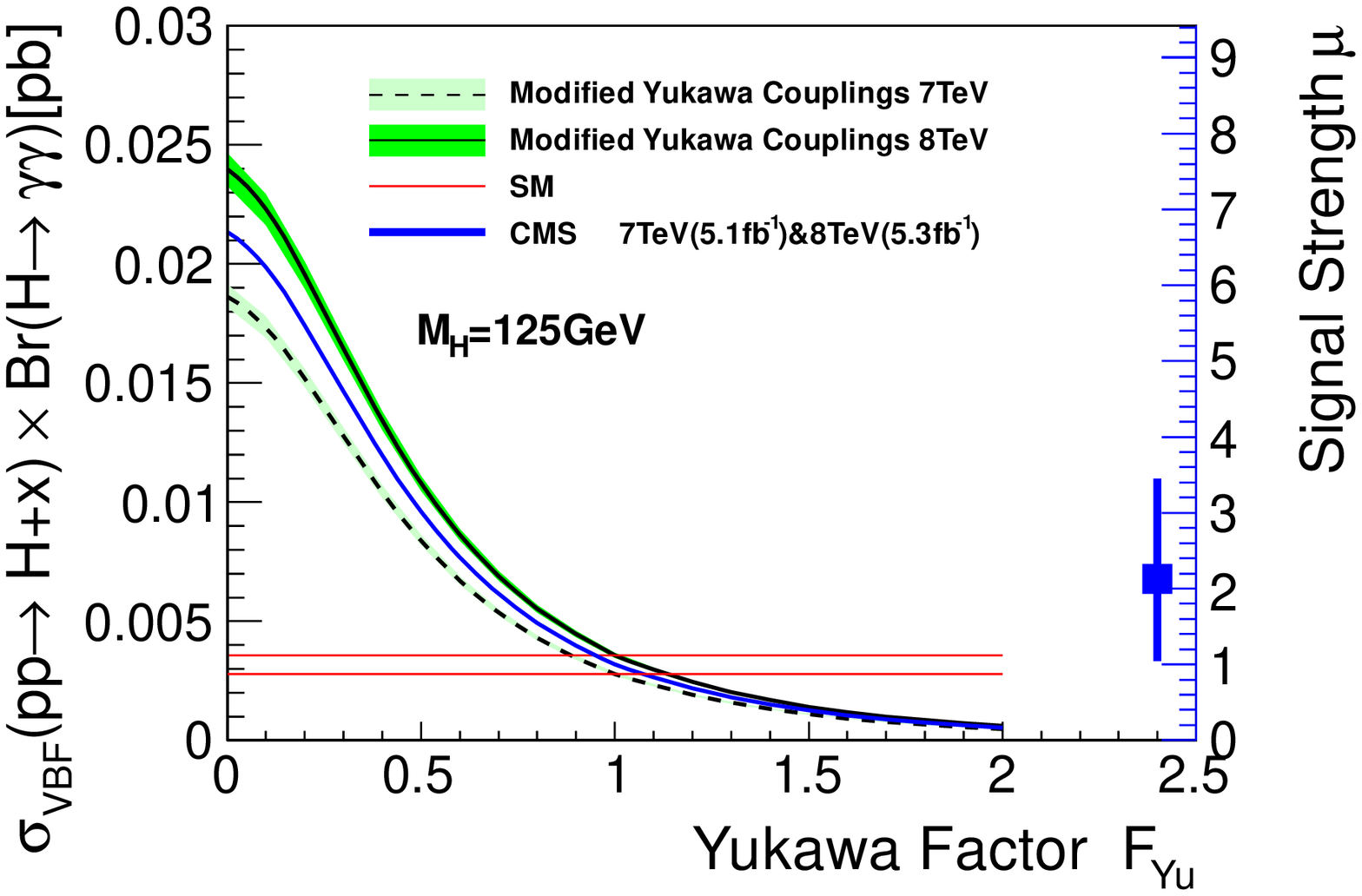}
\includegraphics[width=0.49\textwidth]
{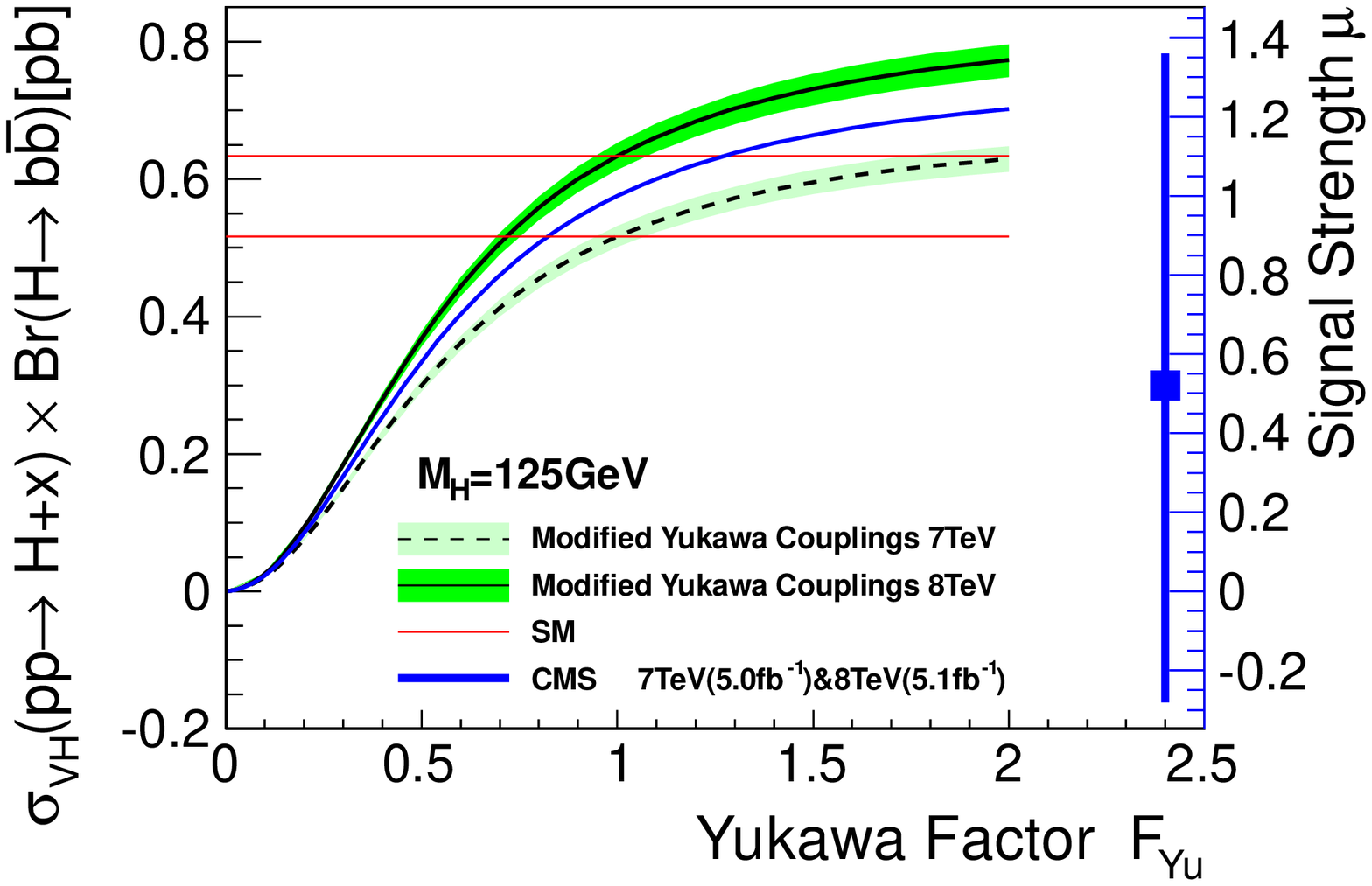}
\end{center}
\caption{\label{VBF78} Higgs observable cross sections versus Yukawa factor $F_{Yu}$ for VBF $H\to \gamma\gamma$/ $V H\to b\bar{b}$.  }
\end{figure}

Figure~\ref{CSgaugebosons78}, \ref{CSff78} and \ref{VBF78} show the comparison of the model predictions and the experimental results for different Higgs production and decay modes. The model predicted observable cross section at 8/7~TeV is drawn as the black/dashed curve with green/light green error band. The SM predicted central values are
shown as the horizontal red line for both 7 and 8~TeV cases. The 7 and 8~TeV curves should be combined together to compare with the experimental signal strength listed in table~\ref{table-expdata78}. This is shown as the magenta/blue curve for the ATLAS/CMS. For simplicity, this combination is done by adding 7~TeV and 8~TeV
predicted values with different weights proportional to their corresponding luminosities. As ATLAS and CMS may have different luminosities for 7 and 8~TeV in the combination, their combined results maybe different and should be compared with their individual experimental results, with different axis-labels, as shown on the right side of each plot.

\begin{figure}[htbp]
\begin{center}
\includegraphics[width=0.49\textwidth]
{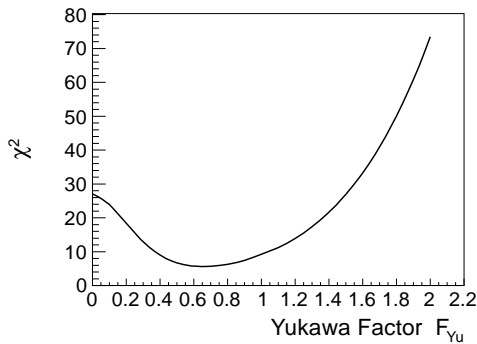}
\end{center}
\caption{\label{chi2-2} The $\chi^2$ fit to find the optimal Yukawa factor $F_{Yu}$. }
\end{figure}

In table~\ref{table-factor78}, we show the allowed $F_{Yu}$ region by different channels for $7\&8$~TeV results, obtained from the above figures. It suggests that an universal suppressed Yukawa coupling factor can increase signal strength easily in VBF, $H\to\gamma\gamma$ channel. However, it maybe difficult to increase the signal strength in the inclusive $H\to\gamma\gamma$ channel. A global $\chi^2$ fit can also be made as shown in equation~(\ref{equation-chi2}). The input data are taken from table~\ref{table-expdata78}. Results are shown in the figure~\ref{chi2-2}, with optimal $F_{Yu}\sim0.65$. This is consistent with the global fit result by CMS collaboration~\cite{CMS:higgsnote}.

\begin{table}[htb]
\centering
\begin{tabular}{|lc|}
\hline
Channels & $F_{Yu}$ region  \\
\hline
$H\to \gamma\gamma$ & 0 \\
$H\to W^+ W^-$ & 0.0-2.0 \\
$H\to Z Z$  & 0.0-2.0 \\
$H\to b\bar{b}$  & 0.0-1.1 \\
$H\to \tau\tau$  & 0.0-0.9 \\
VBF, $H\to\gamma\gamma$ & 0.4-1.0  \\
$VH, H\to b\bar{b}$  & 0.0-2.0 \\
\hline
\end{tabular}
\caption{\label{table-factor78} The allowed $F_{Yu}$ region by different experiment results.  }
\end{table}

\section*{ Acknowledgements}

This research is supported in part
by the Natural Science Foundation of China
under grant numbers 11075003, 10821504, 11075194, and 11135003,
by the Postdoctoral Science Foundation of China under grant
number Y2Y2231B11, and by the DOE grant DE-FG03-95-Er-40917.

\end{document}